\documentclass[aps, prd, onecolumn, tightenlines, notitlepage, superscriptaddress, nofootinbib, preprintnumbers, floatfix,showkeys]{revtex4}

\usepackage[normalem]{ulem}
\usepackage{amstext}
\usepackage{amssymb}
\usepackage{amsmath}
\usepackage{graphicx}
\usepackage{url}
\usepackage{color}
\usepackage{ulem}
\usepackage[utf8]{inputenc}
\pdfoutput=1
\usepackage{textcomp}

\usepackage{epsfig,amsfonts,mathrsfs,amsmath,amssymb,graphicx,color,slashed,multirow}
\usepackage{amsmath,latexsym,amssymb,graphicx,slashed,color,enumerate,url,cancel,gensymb}
\usepackage{textcomp}

\usepackage[x11names]{xcolor}
\usepackage[colorlinks]{hyperref}

\hypersetup{colorlinks,citecolor= nicered,linkcolor= nicered}
\definecolor{nicered}{rgb}{0.7,0.1,0.1}
\definecolor{nicegreen}{rgb}{0.1,0.5,0.1}

\definecolor{vdrgreen}{rgb}{0.0, 0.7, 0.0}

\AtBeginDocument{\hypersetup{citecolor=magenta,linkcolor=blue,urlcolor=nicegreen}}

\begin{document}

\title{{\Large Combined analysis of neutrino decoherence at reactor experiments}}

\author{Andr\'e de Gouv\^ea}\email{degouvea@northwestern.edu}
\affiliation{Northwestern University, Department of Physics \& Astronomy, 2145 Sheridan Road, Evanston, IL 60208, USA}
\author{Valentina De Romeri}\email{deromeri@ific.uv.es}
\affiliation{Institut de F\'{i}sica Corpuscular CSIC/Universitat de Val\`{e}ncia, Parc Cient\'ific de Paterna\\
 C/ Catedr\'atico Jos\'e Beltr\'an, 2 E-46980 Paterna (Valencia) - Spain}
\author{Christoph A. Ternes}\email{ternes@to.infn.it}
\affiliation{INFN, Sezione di Torino, Via P. Giuria 1, I--10125 Torino, Italy}

\begin{abstract}
Reactor experiments are well suited to probe the possible loss of coherence of neutrino oscillations due to wave-packets separation. 
We combine data from the short-baseline experiments Daya Bay and the Reactor Experiment for Neutrino Oscillation (RENO) and from the long baseline reactor experiment KamLAND to obtain the best current limit on the reactor antineutrino wave-packet width, $\sigma > 2.1 \times 10^{-4}$~nm at 90\% CL. 
We also find that the determination of standard oscillation parameters is robust, i.e., it is mostly insensitive to the presence of hypothetical decoherence effects once one combines the results of the different reactor neutrino experiments.
\end{abstract}

\preprint{NUHEP-TH/21-02}
\maketitle

\section{Introduction}

Neutrino oscillations are a consequence of nonzero neutrino masses and the fact that virtually all useful neutrino sources are coherent, i.e., the neutrinos  produced via charged-current weak interactions associated with $\ell_{\alpha}$ charged leptons, $\alpha=e,\mu,\tau$, can be faithfully described as coherent superpositions of neutrinos $\nu_i$ with different masses $m_i$, $i=1,2,3$, weighted by the elements $U_{\alpha i}$ of the leptonic mixing matrix. This is true due in no small part to the fact that neutrino masses are tiny when compared with their laboratory energies. 

While no neutrino source is perfectly coherent, in practice, in the literature, all potential decoherence effects are neglected. This is easy to understand. Pragmatically, real-world considerations, including the finite-size of neutrino sources/detectors and the energy resolution of detecting devices, lead to effects that mimic quantum mechanical decoherence and, for the most part, are quantitatively overwhelming when compared to semi-realistic estimates of the expected fundamental-physics effects. Nonetheless, oscillation experiments can be used to constrain how coherent the different neutrino sources are. 

Nuclear reactors are excellent sources of antineutrinos and reactor antineutrino experiments are  powerful sources of information on the parameters that describe neutrino oscillations. Today, reactor experiments provide the most precise information on the mass-squared difference $\Delta m^2_{21}\equiv m_2^2-m_1^2$ and the magnitude of the $U_{e3}$ ($\sin\theta_{13}$) element of the leptonic mixing matrix\footnote{We use the PDG parameterization for the neutrino oscillation parameters \cite{Zyla:2020zbs}.}. 
They also provide, independent from all other ``types'' of neutrino experiments, valuable information on the magnitude of $\Delta m^2_{31}\equiv m_3^2-m_1^2$ and the magnitude of the product $U_{e1}^*U_{e2}$ ($\propto\sin2\theta_{12}$). Reactor experiments are only {\it in}sensitive to the so-called atmospheric mixing angle $\theta_{23}$ and the CP-odd phase $\delta$. In the near future, the Jiangmen Underground Neutrino Observatory (JUNO) is expected to extract the most precise measurement of $\sin^22\theta_{12}$ and, perhaps, determine the neutrino mass ordering (sign of $\Delta m^2_{31}$)~\cite{Abusleme:2021zrw}. 

In \cite{deGouvea:2020hfl}, we argued that reactor neutrino experiments can also be used to place interesting constraints on how coherent nuclear reactors are as sources of antineutrinos. There, we concentrated on current constraints from the km-baseline experiments Daya Bay, in China,  and the Reactor Experiment for Neutrino Oscillation (RENO), in South Korea, and the near-future JUNO experiment. These experiments are characterized, to a very good approximation, by a single, well-known baseline. In \cite{deGouvea:2020hfl}, we found, for RENO and Daya Bay data, nontrivial correlations between $\sin^2\theta_{13}, |\Delta m^2_{31}|$ and the decoherence parameter $\sigma$. Future JUNO data, in the absence of strong decoherence effects, are expected to significantly weaken these correlations. 

Here, we extend our previous analyses and include data from the Kamioka Liquid Scintillator Antineutrino Detector (KamLAND), which ran for over a decade starting in 2002. Unlike Daya Bay and RENO, KamLAND was sensitive to neutrinos from a large number of nuclear reactor sites located between, roughly, 100~km and 1000~km away and was not characterized by a single baseline. While  Daya Bay and RENO are only sensitive to $\sin^2\theta_{13}$ and $|\Delta m^2_{31}|$, KamLAND is sensitive, given that $|U_{e3}|^2$ is small, only to $\sin^22\theta_{12}$ and $|\Delta m^2_{21}|$ so the data sets, in some sense, complement one another unobstructively. The hypothesis that all reactor experiments are characterized by the same decoherence parameter allows the KamLAND and Daya Bay/RENO data sets to ``inform'' one another in nontrivial ways.

We address the following questions: (i) are the existing reactor neutrino experiments consistent with the hypothesis that nuclear reactors are a source of perfectly coherent antineutrinos (the answer is `yes')?, (ii) if the decoherence parameters of all antineutrinos from nuclear reactors are the same, how well can existing reactor experiments constrain them?, and (iii) if one allows for nontrivial values of the decoherence parameters, how much is the measurement of the different oscillation parameters impacted? 

The paper is structured as follows. In Sec.~\ref{sec:osc}, we discuss neutrino oscillation probabilities including quantum decoherence and matter effects. 
In Sec.~\ref{sec:analysis}, we detail our combined analysis of the existing data from the reactor experiments KamLAND, RENO, and Daya Bay.
In Sec.~\ref{sec:res}, we derive the best current bound on the neutrino wave-packet width and discuss how the presence of neutrino decoherence may affect the extraction of standard oscillation parameters.
We draw our conclusions in Sec.~\ref{sec:conc}.

\section{Neutrino oscillations including decoherence in matter}
\label{sec:osc}
Reactor neutrino experiments detect the flux of electron antineutrinos produced in nuclear reactor cores through the inverse beta-decay process, $\bar{\nu}_e + p \rightarrow e^+ + n$. Keeping in mind uncertainties on the antineutrino flux produced in the nuclear fission processes, these experiments can measure the $\bar{\nu}_e$ survival probability $P(\bar{\nu}_e \to \bar {\nu}_e)$, which depends on the energy of the antineutrinos $E$ and the baseline $L$, the distance between the source and the detector.

While decoherence effects may stem from several different physical origins~\cite{Kiers:1995zj,Ohlsson:2000mj,Beuthe:2001rc,Beuthe:2002ej,Giunti:2003ax,Blennow:2005yk,Farzan:2008eg,Kayser:2010pr,Naumov:2010um,Naumov:2013uia,Jones:2014sfa,Akhmedov:2019iyt,Grimus:2019hlq,Naumov:2020yyv},  here we focus on the possible loss of flavor-coherence of the neutrino beam that grows with the baseline (often referred to as wave-packet separation) and is parameterized through the damping parameters
\begin{equation}
\xi_{jk}(L,E)=\bigg( \frac{L}{L^{\rm coh}_{jk}} \bigg)^2  ~~{\rm with}~~ j,k = 1,2,3.
\label{eq:dec_fac}
\end{equation}
We refer to \cite{deGouvea:2020hfl} for a more careful discussion of these parameters and their effects. 
If no loss of coherence occurs during neutrino propagation, $\xi_{jk} (= \xi_{kj})=0$. 
We further define the coherence lengths as~\cite{Giunti:1991sx,Beuthe:2002ej,Kayser:2010pr,deGouvea:2020hfl}
\begin{equation}
  L^{\rm coh}_{jk} = \frac{4 \sqrt{2} E^2}{|\Delta m_{jk}^2|} \sigma\, ,
  \label{eq:Lcoh}
\end{equation}
which depend on the neutrino energy and the mass-squared differences.
We assume all decoherence effects to be encoded in a single parameter $\sigma$, which can be interpreted as the width of the neutrino wave-packet and has dimensions of length.
Estimates for the typical value of $\sigma$  depend on the physics responsible for neutrino production and vary by orders of magnitude. This physics contains several distance-scales, many of which potentially inform $\sigma$. For instance, distance scales associated with electron antineutrinos detected via inverse beta-decay may include: i) the typical size of the beta-decaying nuclei ($\sim 10^{-5}$ nm) ii) typical interatomic spacing that characterizes the fuel ($\sim 0.1 -1$ nm for Uranium) iii) the inverse of the antineutrino energy ($\sim 10^{-4}$ nm).
Without adding to the discussion of estimating $\sigma$, we choose to stay agnostic and consider $\sigma$ as a free parameter, generically assuming that it will depend on the features of neutrino production and detection.

In the presence of decoherence effects, the density matrix $\rho_{jk}$ describing the flavor content of the reactor antineutrinos produced in nuclear power plants as a function of $L$ and $E$ is 
\begin{equation}
\tilde{\rho}_{jk}(L,E)  = \tilde{U}_{ej}^*\tilde{U}_{ek}\exp[-i\tilde{\Delta}_{jk}]\exp[-\tilde{\xi}_{jk}(L,E)]\,,
\end{equation}
with 
\begin{equation}
\tilde{\Delta}_{jk}\equiv 2\pi\frac{L}{\tilde{L}^{\rm osc}_{jk}} \equiv \frac{\Delta\tilde{m}^2_{jk}L}{2E}\,,
\end{equation}
where we have included matter effects (assuming the antineutrinos propagate through a medium with constant density) by substituting all the quantities in vacuum with the corresponding well-known effective matter-quantities~\cite{Giunti:1991sx}. The tilde in the variables defined above denotes that a quantity is affected by matter effects.
In particular, $\tilde{\xi}_{jk}$ is as defined in Eq.~\eqref{eq:dec_fac} but with the mass-squared difference replaced by its matter counterpart.
The electron antineutrino survival probability in presence of decoherence effects (for a constant matter profile) is given by the $ee$ diagonal element of the density matrix:
\begin{align}
  P^\text{dec}(\overline{\nu}_e \to \overline{\nu}_e) =
    \sum_{j,k} |\tilde{U}_{e j}|^2 |\tilde{U}_{e k}|^2 \,
        \exp[
          - i \tilde{\Delta}_{jk}
          - \tilde{\xi}_{jk}] \,.
          \label{eq:enuprob}
\end{align}
As expected, we recover the standard oscillation probability when the damping factors $\tilde{\xi}_{jk} \rightarrow 0$ or, equivalently, when  $\tilde{L}^{\rm coh}_{jk} \rightarrow \infty$ ($\sigma \rightarrow \infty$).

We compute the oscillation parameters in matter with the help of the parameterization discussed in Ref.~\cite{Denton:2016wmg}, which has been shown to be among the most precise and most efficient ones~\cite{Parke:2019vbs}. The matter effects, in principle, render the current reactor neutrino experiments sensitive to the neutrino mass ordering. In practice, however, the sensitivity is completely negligible.  
Current analyses of the world neutrino data slightly prefer the normal mass ordering~\cite{deSalas:2020pgw,Capozzi:2017ipn,Esteban:2020cvm,Kelly:2020fkv} so, for concreteness, we assume it to be normal (i.e., $\Delta m^2_{31}>0$). Had we performed the analysis assuming the inverted ordering, we would have obtained the same results. A little more quantitatively, on the Earth's crust, the matter potential $\sqrt{2}G_Fn_e\sim 10^{-7}$~eV$^2$/MeV, where $n_e$ is the electron number density. This is to be compared with $\Delta m^2_{ij}/2E$. For reactor antineutrinos,  $\Delta m^2_{21}/2E\gtrsim 5\times 10^{-6}$~eV$^2$/MeV, twenty times larger ($|\Delta m^2_{31}|$, of course, is thirty times larger than $\Delta m^2_{21}$). Hence, matter effects will impact oscillations at, very roughly, the 5\% level in experiments, like KamLAND, sensitive to $\Delta m^2_{21}$. This is especially true of KamLAND, where very long baselines ($L={\cal O}$(1000~km)) are relevant, keeping in mind that $\sqrt{2}G_Fn_e \times1000~\rm km\sim 0.3$. 

\begin{figure}
\centering
\includegraphics[width=0.49\textwidth]{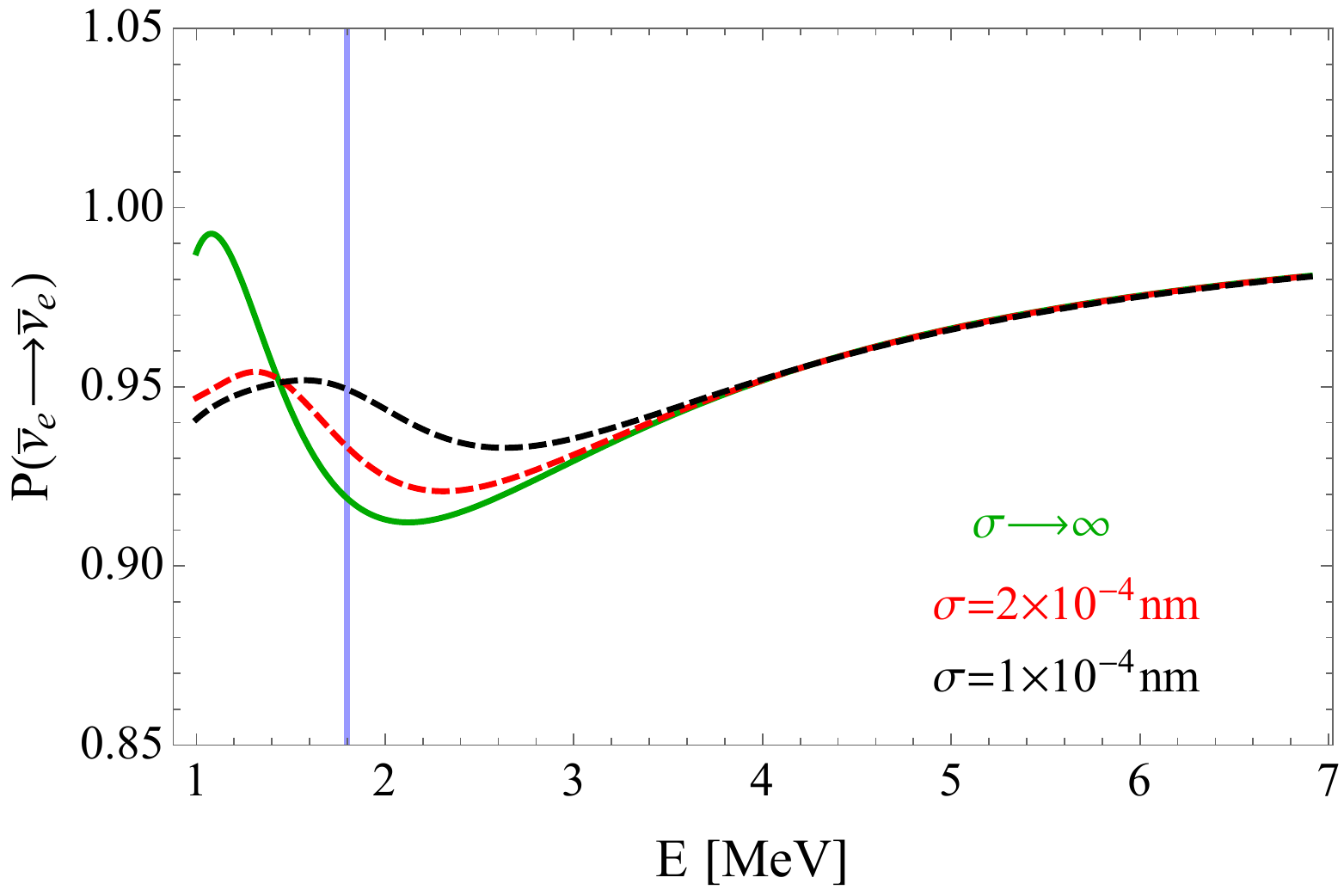}
\includegraphics[width=0.49\textwidth]{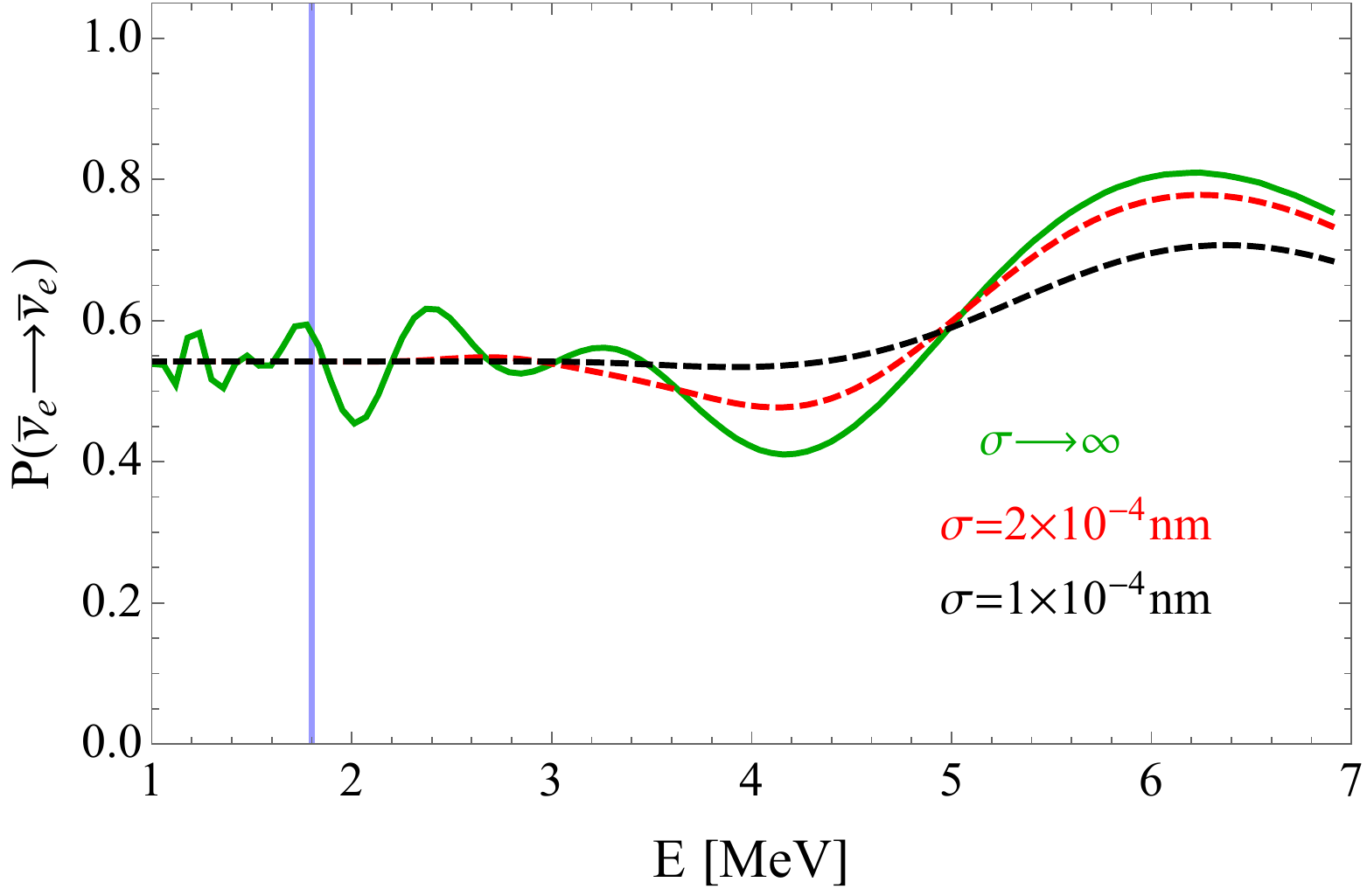}
\caption{Left: the electron antineutrino oscillation probability as a function of the neutrino energy for $L=1$~km, representative of short-baseline experiments like RENO and Daya Bay. Right: the average electron antineutrino oscillation probability at KamLAND. In both panels, the colors correspond to different values of the decoherence parameter $\sigma$. The light blue, vertical line sets the threshold for inverse beta-decay detection. All other oscillation parameters are fixed at their current best-fit values, see \cite{deSalas:2020pgw}, and we assume that the mass ordering is normal.}
\label{fig:osc_prob_Kaml}
\end{figure}

In order to illustrate the impact of decoherence on reactor antineutrino oscillations at the different reactor experiments, we depict in Fig.~\ref{fig:osc_prob_Kaml} (left) the expected electron antineutrinos survival probability for $L=1$~km, representative of the baselines of the Daya Bay and JUNO experiment, while in Fig.~\ref{fig:osc_prob_Kaml} (right) we depicted the average electron antineutrino survival probability at KamLAND. In the latter, we perform a weighted average taking into account the different baselines and reactor power outputs. We fix all standard oscillation parameters to the current best-fit values extracted in Ref.~\cite{deSalas:2020pgw}.
The green, solid curve corresponds to the standard neutrino oscillation scenario without decoherence, while the red and black dashed ones are obtained assuming non-trivial decoherence effects associated to $\sigma = 2 \times 10^{-4}$~nm and $\sigma = 1 \times 10^{-4}$~nm, respectively. These values are consistent with the lower limits obtained in the current work (see~Sec.~\ref{sec:res}) and in our previous analysis of RENO + Daya Bay data~\cite{deGouvea:2020hfl}, respectively. The impact of a finite value for $\sigma$ is clear in Fig.~\ref{fig:osc_prob_Kaml}. Decoherence ``erases'' the oscillatory behavior of the survival probability and its impact is more pronounced at relatively smaller energies. 

\section{Data analysis}
\label{sec:analysis}

We analyze data from three different reactor experiments. 
RENO and Daya Bay are reactor neutrino experiments in South Korea and China, respectively, that measure the flux of antineutrinos from nuclear reactors at baselines $L\sim 100$~m and $L\sim 1$~km, using information from both the near and far detectors to measure the electron antineutrino survival probability, Eq.~(\ref{eq:enuprob}). We rely on the experimental data presented in Refs.~\cite{Adey:2018zwh} and~\cite{jonghee_yoo_2020_4123573}.
The procedure followed in this paper builds upon our previous analysis\footnote{See also Ref.~\cite{An:2016pvi} for an analysis with a smaller Daya Bay dataset.} described in Ref.~\cite{deGouvea:2020hfl}, with the only difference that here we also allow for the values of $\Delta m_{21}^2$ and $\theta_{12}$ to vary in the fit.
We refer the reader to our previous work~\cite{deGouvea:2020hfl} for more technical details and for the definition of the $\chi^2$ functions $\chi^2_\text{RENO}$ and $\chi^2_\text{DB}$.

KamLAND was a long-baseline reactor experiment, placed at the site of the former Kamiokande experiment, measuring antineutrinos from more than 50 reactor cores and distances ranging from $\sim 100$~km to $\sim 1000$~km. 
As discussed earlier, because of the long baselines, matter effects are not negligible at KamLAND, and are included in our calculations, as detailed in Sec.~\ref{sec:osc}.
We follow the analysis and include the data presented in Refs.~\cite{Gando:2010aa,kamland_web}. 
The $\chi^2$ function for KamLAND is

\begin{equation}
 \chi^2_\text{KL}(\vec{p}) = \min_{\vec{\alpha}}\left\{\sum_{i=1}^{N_\text{KL}}\left(\frac{N_{\text{dat},i} - N_{\text{exp},i}(\vec{p},\vec{\alpha})}{\sigma^\text{KL}_i}\right)^2 + \frac{(N^{\text{tot}}_{\text{dat}} - N^{\text{tot}}_{\text{exp}}(\vec{p},\vec{\alpha}))^2}{N^{\text{tot}}_{\text{dat}}}
+ 
 \sum_k \left(\frac{\alpha_k - \mu_k}{\sigma_k} \right)^2\right\}\,,
\end{equation}
where the index $i$ runs over the energy bins.
$N_{\text{dat},i}$ are the observed event numbers, while $N_{\text{exp},i}(\vec{p},\vec{\alpha})$ are the expected event numbers for a given set of oscillation parameters $\vec{p}$. 
Following the collaboration approach, we include a penalty term on the total number of events.
The last term contains penalty factors for all of the systematic uncertainties $\alpha_k$ with expectation value $\mu_k$ and standard deviation $\sigma_k$. 
We include several sources of systematic uncertainties, accounting for reactor uncertainties (normalizations related to the different reactors and an uncorrelated shape error) and detector uncertainties (detection efficiency and energy scale).
We use GLoBES~\cite{Huber:2004ka,Huber:2007ji} to compute the event numbers and to perform the statistical analysis.
The reactor fluxes are parameterized as in Ref.~\cite{Huber:2004xh} and the inverse beta-decay cross section is taken from Ref.~\cite{Vogel:1999zy}. 
After analyzing the data from each experiment independently, we also combine the three data sets using

\begin{equation}
 \chi^2_\text{COMB}(\vec{p}) = \chi^2_\text{RENO}(\vec{p}) + \chi^2_\text{DB}(\vec{p}) +
 \chi^2_\text{KL}(\vec{p}) \,.
\end{equation}

In order to validate our treatment of the data sets, we first assume a perfectly coherent source and compare our results to those published by the experimental collaborations~\cite{Adey:2018zwh,jonghee_yoo_2020_4123573,Gando:2010aa}.
Hence, we first consider the case $\vec{p} = (\Delta m_{31}^2, \theta_{13}, \Delta m_{21}^2, \theta_{12})$.
Next, to explore the impact of decoherence effects, we allow for the possibility of a finite wave-packet width $\sigma$ and consider an extended set of parameters, $\vec{p} = (\Delta m_{31}^2, \theta_{13}, \Delta m_{21}^2, \theta_{12}, \sigma)$.
%

\section{Results}
\label{sec:res}
\begin{figure}
\centering
\includegraphics[width=0.9\textwidth]{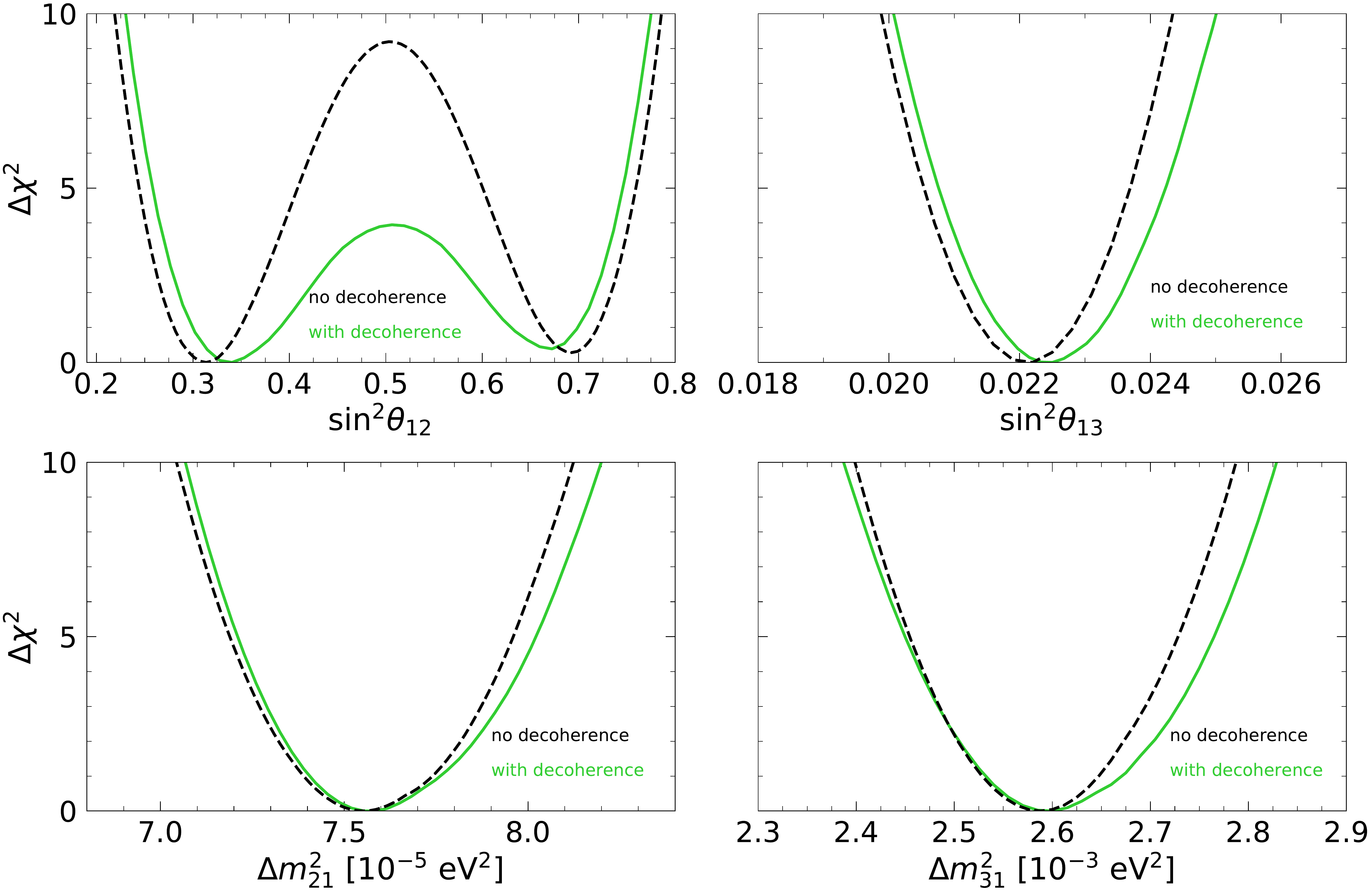}
\caption{
The reduced $\chi^2$ profiles as a function of the standard neutrino oscillation parameters, obtained from the combined analysis RENO + Daya Bay + KamLAND including decoherence effects (green, solid) and assuming a perfectly coherent source (black, dashed).  The profiles are calculated with respect to the global minimum for each case.}
\label{fig:chi2_profiles}
\end{figure}

We start by comparing the $\Delta \chi^2$ profiles obtained from the combination of all experiments in the presence of decoherence effects with those obtained assuming a perfectly coherent source. The reduced $\chi^2$ profiles for all four accessible oscillation parameters are depicted in Fig.~\ref{fig:chi2_profiles}.  
In all plots, we marginalize over all absent parameters, including $\sigma$ when decoherence effects are allowed in the fit. The best-fit values for the standard neutrino oscillation parameters are not strongly affected by the possible loss of coherence. In particular, the best fit values of the mass squared differences $\Delta m^2_{21}$ and $\Delta m^2_{31}$ are identical regardless of whether decoherence effects are allowed in the fit.  Remember that we assume  $\Delta m^2_{31}$ to be positive. The reduced-$\chi^2$ functions are somewhat shallower when decoherence effects are allowed in the fit, as expected.
The best-fit values for $\sin^2\theta_{12}$ and $\sin^2\theta_{13}$ are instead slightly altered, moving towards slightly larger values of $\sin^22\theta_{12,13}$. Also in this case, and as expected, the reduced-$\chi^2$ functions are somewhat shallower when decoherence effects are allowed in the fit. Fig.~\ref{fig:chi2_profiles} (top,left) is a reminder that reactor antineutrino experiments are mostly sensitive to $\sin^22\theta_{12}$ and cannot really distinguish $\theta_{12}$ from $\pi/2-\theta_{12}$. This degeneracy is resolved by solar neutrino data \cite{PhysRevD.54.2048,deGouvea:1999xe,deGouvea:2000pqg}, which also provides a more precise determination of $\sin^2\theta_{21}$ relative to reactor experiments\footnote{There is a similar degeneracy when it comes to the determination of $\sin^2\theta_{13}$; values close to one, however, are very safely ruled out by the remainder of the world's neutrino oscillation data.}.  Note that the small asymmetry in $\sin^2\theta_{12}$ is due to the inclusion of matter effects in our analysis.
In summary, the determination of the standard oscillation parameters is not substantially impacted by the possible loss of coherence of neutrino oscillations due to neutrino wave-packet separation.

\begin{figure}
\centering
\includegraphics[width=0.45\textwidth]{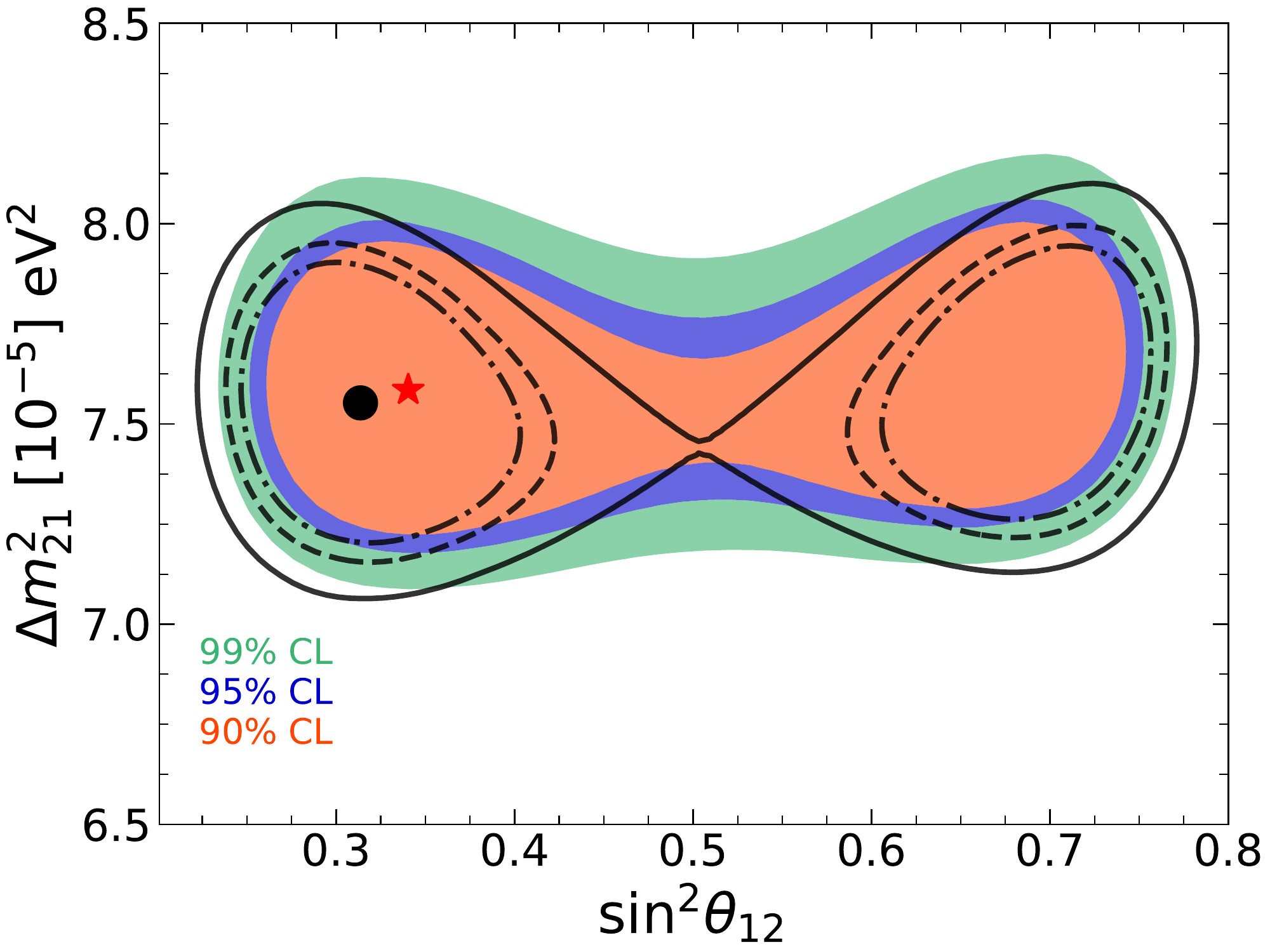}
\includegraphics[width=0.45\textwidth]{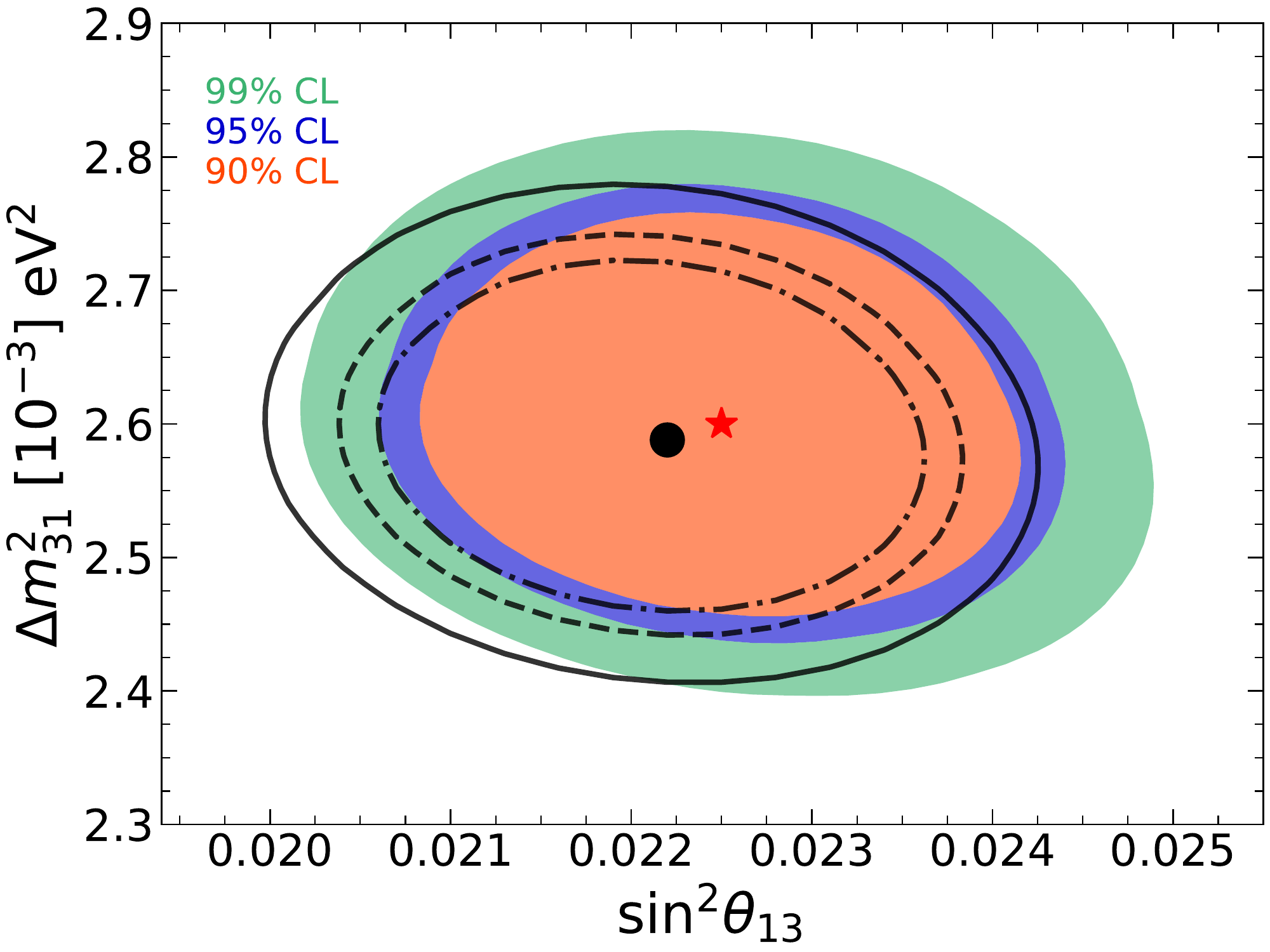}
\caption{
90, 95 and 99\% CL (2 d.o.f.) allowed regions in the $\sin^2\theta_{12}$-$\Delta m_{21}^2$ (left) and $\sin^2\theta_{13}$-$\Delta m_{31}^2$ (right) planes from our combined analysis of RENO + Daya Bay + KamLAND data including decoherence (filled regions, red stars) and assuming a perfectly coherent source (black empty contours, black dots).}
\label{fig:sq13_dm31_s12_dm21}
\end{figure}

Fig.~\ref{fig:sq13_dm31_s12_dm21} depicts the $\sin^2\theta_{12}$-$\Delta m_{21}^2$ (left) and $\sin^2\theta_{13}$-$\Delta m_{31}^2$ (right) regions of parameter space consistent with the combined data sets (filled regions in orange at 90\% CL, blue at 95\% CL, green at 99\% CL). In all plots, we marginalize over all absent parameters, including $\sigma$ when decoherence effects are allowed in the fit.
The figure also depicts the allowed contours corresponding to the analysis performed assuming a perfectly coherent source (black empty curves, dot-dashed at 90\% CL, dashed at 95\% CL, solid at 99\% CL).  
The best-fit points from the standard analyses are indicated with black dots while the best-fit value from the analyses including nontrivial $\sigma$ are marked with red stars.
Decoherence effects shift the allowed regions towards larger values of $\sin^22\theta_{12,13}$ and lead to larger (i.e., larger area) allowed regions at the same CL. In the left panel, nonetheless, note that the allowed region for $\sin^2\theta_{12}$ becomes a bit smaller when including decoherence effects (see also Fig.~\ref{fig:chi2_profiles} (top,left)). 

In~\cite{deGouvea:2020hfl}, relying only on RENO + Daya Bay data, we found a relatively stronger correlation among the parameters, with larger values of  $\sin^2\theta_{13}$ or smaller values of $\Delta m_{31}^2$ compensating for, respectively, the decoherence effects of flattening the oscillation maximum due to the presence of $\sigma$ or shifting the position of the first oscillation maximum to smaller $L/E$ values.
Here, the combination of data from short-baseline experiments together with those from KamLAND significantly reduces the allowed region in the $\sin^2\theta_{13}$-$\Delta m_{31}^2$ (compare  Fig.~\ref{fig:sq13_dm31_s12_dm21} (right) with Fig.~2 of Ref.~\cite{deGouvea:2020hfl}). 
This is a consequence of the fact that KamLAND is more sensitive to nontrivial $\sigma$ effects that RENO and Daya Bay, as we turn to momentarily. Data from KamLAND, it turns out, excludes the relatively small values of $\sigma$ that allow for the stronger correlations we reported in~\cite{deGouvea:2020hfl}. 

When it comes to the so-called solar parameters, we do not observe significant correlations between $\sigma$ and $\Delta m_{21}^2$. This translates into a lack of unexpected correlations in Fig.~\ref{fig:sq13_dm31_s12_dm21} (left).
The main reason for this is that KamLAND observes both an oscillation maximum and a minimum, reducing the ability of compensating for decoherence effects by changing the value of $\Delta m^2_{21}$. Values of $\sigma$ that are small enough to ``flatten'' the minimum (which flattens faster than the maximum, as illustrated in Fig.~\ref{fig:osc_prob_Kaml}) are excluded, and we end up with a robust measurement of $\Delta m_{21}^2$. This is unlike the situation at Daya Bay and RENO, which instead observe only the first oscillation minimum. 

\begin{figure}
\centering
\includegraphics[width=0.5\textwidth]{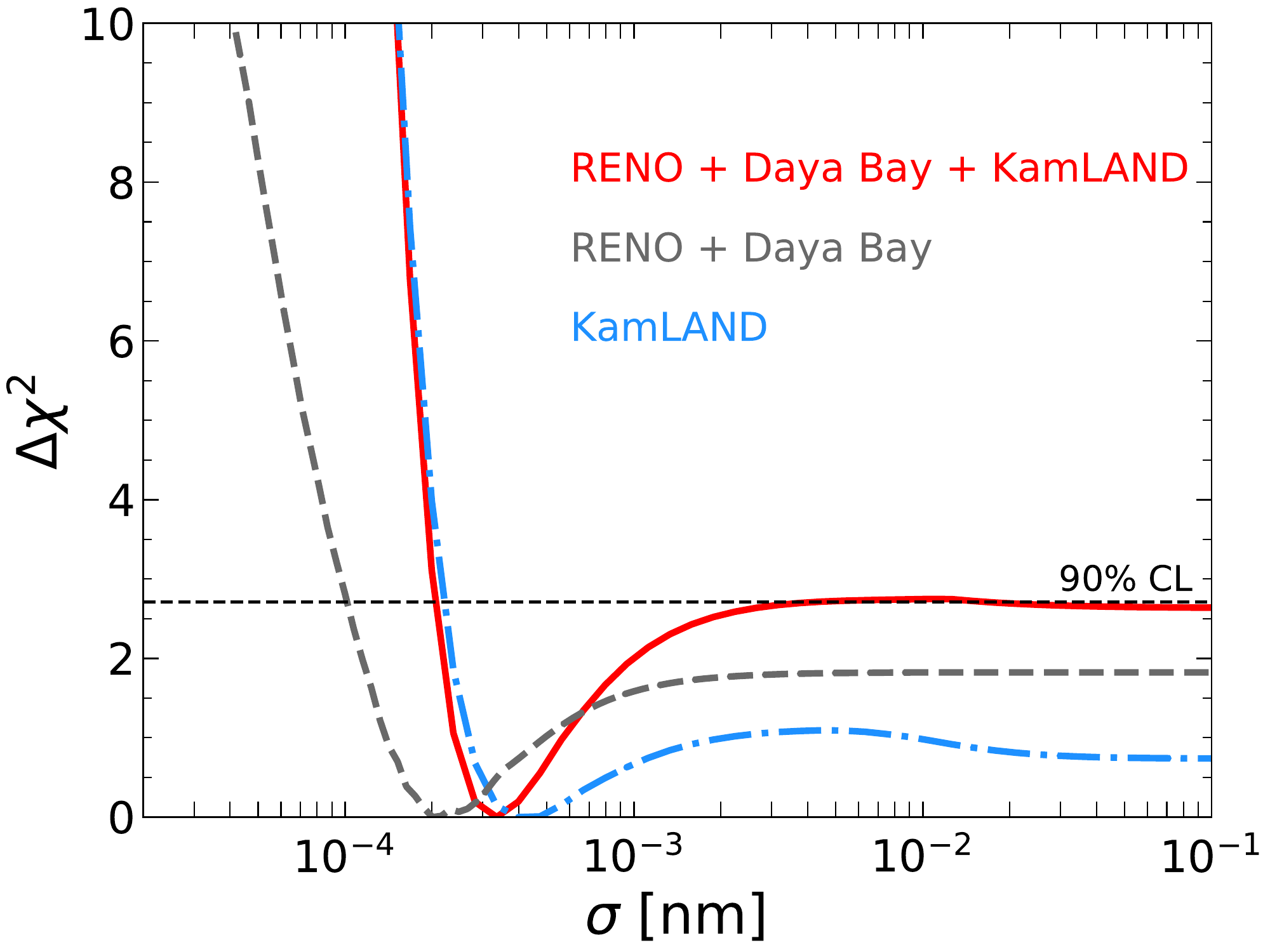}
\caption{
The reduced $\chi^2$ as a function of $\sigma$ relative to its minimum value, obtained from the combined analysis of RENO, Daya Bay and KamLAND (red, solid) and from the combined analysis of only short-baseline experiments (grey, dashed).}
\label{fig:sigma_profile}
\end{figure}

Fig.~\ref{fig:sigma_profile} depicts the reduced $\chi^2$ as a function of $\sigma$, relative to the minimum value. 
Marginalizing over the standard oscillation parameters, we obtain the following best-fit value for the reactor-antineutrino-wave-packet width: $\sigma=3.35\times 10^{-4}$~nm. 
The no-decoherence hypothesis, $\sigma\to\infty$,  however, is safely allowed at 90\% CL and we can infer a lower bound on $\sigma$:  $\sigma > 2.08\times 10^{-4}$~nm at 90\% CL. This is stronger by a factor 2 relative to the previous lower bound $\sigma > 1.02\times10^{-4}$~nm~\cite{deGouvea:2020hfl}, obtained by combining data only from RENO and Daya Bay.
As mentioned earlier, this explains why we do not observe strong correlations in the $\sigma$-$\sin^2\theta_{13}$-$\Delta m_{31}^2$ parameter ``volume.'' While KamLAND data are virtually blind to $\sin^2\theta_{13}$ and $\Delta m_{31}^2$, they translated into a relatively stronger bound on $\sigma$ which in turn breaks the degeneracy observed in Ref.~\cite{deGouvea:2020hfl}. KamLAND data also exclude, at the 90\%~CL, the value of $\sigma$ preferred by the data from Daya Bay and RENO.

\section{Conclusions}
\label{sec:conc}

We computed the effects of wave-packet-separation decoherence on the current data on the oscillations of reactor antineutrinos, obtained for $L\sim$~1~km (Daya Bay and RENO) and for $L\sim$~100+~km (KamLAND). We found that the current data can exclude wave-packet sizes $\sigma<2.08\times 10^{-4}$~nm at 90\% CL, assuming that neutrinos from all nuclear-reactor cores can be characterized by the same $\sigma$. We also studied the impact of allowing for arbitrary values of $\sigma$ when measuring $\sin^2\theta_{12}$, $\sin^2\theta_{13}$, $\Delta m^2_{21}$, and $|\Delta m^2_{31}|$. We found that, given the existing reactor data, these measurements are robust, i.e., regardless of whether nontrivial $\sigma$ values are allowed in the fit, the extracted best-fit values and error bars are approximately the same and no unusual correlations are induced by allowing for  nontrivial $\sigma$ values.

We found that KamLAND data are more sensitive to decoherence effects than those of Daya Bay and RENO combined. This is not a trivial statement. Daya Bay and RENO have accumulated more statistics. Furthermore, KamLAND ``sees'' neutrinos from a plurality of nuclear cores and averaging-out effects, which tend to mimic those of decoherence, are very significant. It turns out, however, that the access to very long baselines, the relatively large value of $\sin^22\theta_{12}$, and the fact that KamLAND ``sees'' both oscillation maxima and minima leads to stronger sensitivity. In the next few years, we expect an order-of-magnitude better sensitivity from the JUNO experiment \cite{deGouvea:2020hfl, Cheng:2020jje}.

As in our previous publication \cite{deGouvea:2020hfl}, here we also choose not to add to the very interesting but subtle discussion of expectations for  $\sigma$ given antineutrinos produced in nuclear-reactor cores. We reiterate that naive estimates are safely larger than the experimental bound obtained here. Nonetheless, we find it is important to test the hypothesis that nuclear reactors are, for modern practical applications, a coherent source of antineutrinos, to probe how large decoherence effects could be, and to understand how these might impact our ability to measure fundamental physics parameters with reactor neutrino oscillation experiments.

\section*{Acknowledgments}

We thank the participants and organisers of ``The Magnificent CE$\nu$NS workshop 2020" for interesting conversations and comments that inspired this work.
The work of AdG is supported in part by the DOE Office of Science award \#DE-SC0010143.
VDR acknowledges financial support by the SEJI/2020/016 grant (project ``Les Fosques'') funded by
Generalitat Valenciana, by the Universitat de Val\`encia through the sub-programme “ATRACCI\'O DE TALENT 2019” and partial support by the Spanish grant FPA2017-85216-P. 
CAT is supported by the research grant ``The Dark Universe: A Synergic Multimessenger Approach'' number 2017X7X85K under the program ``PRIN 2017'' funded by the Ministero dell'Istruzione, Universit\`a e della Ricerca (MIUR). 

\bibliographystyle{utphys}
\providecommand{\href}[2]{#2}\begingroup\raggedright\endgroup

\end{document}